\documentclass{elsart}
\makeatletter
\@addtoreset{equation}{section}
\@addtoreset{table}{section}
\makeatother
\newcommand{\eref}[1]{(\ref{#1})}

\def\I{{\rm i}}
\def\D{{\rm d}}

\usepackage{amsmath,amssymb,array}
\begin{document}

\begin{frontmatter}

\title{Supersymmetry in the Non-Commutative Plane}

\author[Talca]{Luc Lapointe},
\ead{lapointe@inst-mat.utalca.cl}
\author[CRM]{Hideaki Ujino\thanksref{GCT}} and 
\thanks[GCT]{On leave of absence from
Gunma National College of Technology, Maebashi, Gunma 371--8530, Japan,
{\tt ujino@nat.gunma-ct.ac.jp}}
\ead{ujino@crm.umontreal.ca}
\author[McGill]{Luc Vinet}
\ead{luc.vinet@mcgill.ca}

\address[Talca]{Instituto de Matem\'atica y F\'{\i}sica, 
Universidad de Talca,\\
Casilla 747, Talca, Chile}
\address[CRM]{Centre de recherches math\'ematiques, 
Universit\'e de Montr\'eal,\\
Montreal, Quebec, Canada, H3C 3J7}
\address[McGill]{Department of Mathematics and Statistics 
and Department of Physics,\\
McGill University,
Montreal, Quebec, Canada, H3A 2J5}

\maketitle

\begin{abstract}
The supersymmetric extension of a model introduced by
Lukierski, Stichel and Zakrewski in the non-commutative plane
is studied. 
The Noether charges associated to the symmetries are determined.
Their Poisson algebra is investigated in the Ostrogradski--Dirac formalism 
for constrained Hamiltonian systems. 
It is shown to provide a supersymmetric generalization of the Galilei algebra
with a two-dimensional central extension.
\end{abstract}

\end{frontmatter}

\section{Introduction}
There has been an increased interest lately in the study of physics
in non-commutative space-time. This stems, in particular, from advances in
string theory~\cite{SW} and from the Connes program~\cite{Connes1,Connes2}
(see also~\cite{Jackiw}). In this context, two models~\cite{DH1,LSZ}
with interesting features were recently introduced 
in the non-commutative plane in independent and different ways.
In the two cases, which have been shown to be related with 
each other~\cite{DH2},\footnote{Quite recently, a slight difference 
between the models in refs.~\cite{DH1,LSZ} was reported~\cite{HP2}.}
the non-commutativity of the space coordinates is intimately
related to the invariance of the model under the Galilei group with
a two-dimensional central extension. 
While the dynamics is described using coadjoint orbits and 
canonical symplectic structures in ref.~\cite{DH1}, 
the Lagrangian picture is used in ref.~\cite{LSZ}.  In this paper,
we shall focus exclusively on the latter description, which
posits a non-relativistic classical model in two dimensions
described by the Lagrangian 
\begin{equation}
  L_{\rm b} := \dfrac{1}{2}m\dot{x}_i^2-k\epsilon_{ij}\dot{x}_i 
  \ddot{x}_j, \quad i,j=1,2,
  \label{eq:LSZ}
\end{equation}
where $\epsilon_{ij}$ is the Levi--Civita antisymmetric metric~\cite{LSZ},
and where, as will be the case throughout the paper, 
the Einstein convention on the summation of repeated indices is employed. 
The model~\eref{eq:LSZ} was shown to have 
the (2+1)-Galilean symmetry~\cite{LL} with a two-dimensional central
extension parametrized by
the mass $m$ and the coupling parameter $k$.

We study in this paper a supersymmetrized version of 
the model~\eref{eq:LSZ}, which has been also introduced in 
ref.~\cite{LSZ2}.\footnote{The authors are thankful to
Prof.~P.~C.~Stichel for drawing their attention to this reference.}
 In addition to the intrinsic interest of the
generalized model, an additional motivation is the exploration
of the supersymmetric enlargement of the Galilei algebra with a
two-dimensional central extension. We shall also identify the presence
of the higher conformal and superconformal symmetries in the original
and in the supersymmetric models.

The paper is organized as follows.
In section~\ref{sec:model}, we introduce the supersymmetric model.
In section~\ref{sec:eom}, the equations of motion and the canonical 
structure of the supersymmetric model are presented through 
the Ostrogradski--Dirac formalism.
In section~\ref{sec:lpa}, we obtain the Noether charges associated with
the symmetries and investigate the Poisson algebra that they generate.
The final section includes a summary and concluding remarks.

\section{The supersymmetric model and its symmetries}
\label{sec:model}

We shall consider a generalization of the
Lagrangian \eref{eq:LSZ} involving a two-dimensional
free-fermion term:
\[
  L
  =L_{\rm b}+\dfrac{\I}{2}\xi_i\dot{\xi}_i+\cdots,
  \quad \xi_i,\  i=1,2:\  \text{Grassmannian}
\]
and supplemented by additional terms so that $L$ be
invariant under the infinitesimal supersymmetric 
transformation,
\begin{equation}
  \begin{split}
  & \delta_{Q}x_i:=\I\alpha\xi_i,\ \delta_{Q}\xi_i:=-m\alpha \dot{x}_i \\
  & \alpha:\ \text{an infinitesimal Grassmannian parameter}
  \end{split}
  \label{eq:supert}
\end{equation}
up to a total time-derivative, 
$\delta_Q L=\frac{\D \Lambda_Q}{\D t}$.
It is straightforward to check that the following Lagrangian
\begin{equation}
  L=L_{\rm b}+L_{\rm f}
  :=\dfrac{1}{2}m\dot{x}_i^2-k\epsilon_{ij}\dot{x}_i\ddot{x}_j
  +\dfrac{\I}{2}\xi_i\dot{\xi}_i+\dfrac{\I k}{m}\epsilon_{ij}
  \dot{\xi}_i\dot{\xi}_j
  \label{eq:SLSZ}
\end{equation}
remains invariant under the infinitesimal supersymmetric 
transformation \eref{eq:supert} up to a total time-derivative.%
\footnote{Equation (14) in ref.~\cite{LSZ2} reads as
$L_{\rm SUSY}^{(0)}\sim L_{\rm b}+mL_{\rm f}$, which is 
essentially the same as eq.~\eref{eq:SLSZ}.
The slight difference in the Lagrangians comes from the non-essential
difference in the definitions of the supersymmetric transformation
(given in this article by eq.~\eref{eq:supert} and 
in ref.~\cite{LSZ2} by $\delta_{\rm Q}\xi_i=-\alpha\dot{x}_i$).}
To be more specific, we have
\begin{equation}
  \delta_Q L
  =\dfrac{\D \Lambda_Q}{\D t},
  \quad \Lambda_{Q}:=\alpha\Bigl(\dfrac{1}{2}\I m\dot{x}_i\xi_i-\I k 
  \epsilon_{ij}\dot{x}_i\dot{\xi}_j\Bigr).
  \label{eq:LambdaQ}
\end{equation}
We shall refer to the system described by the Lagrangian~\eref{eq:SLSZ}
as the sLSZ model.

Extending the symmetry analysis of $L_{\rm b}$~\cite{LSZ,LL} to
the system containing the Grassmannian variables~\cite{C1,C2}, 
we observe that the sLSZ model exhibits Galilean supersymmetry.
The corresponding transformations take the form,
\begin{equation}
  \begin{split}
    & \delta_{\rm r}x_i:= -\epsilon_{ij}x_j\phi_{\rm b},\ 
    \delta_{\rm r}\xi_i:=-\epsilon_{ij}\xi_j\phi_{\rm f},\ 
    \delta_{\rm G}x_i:=v_i t \\ 
    & \delta_{\rm t}x_i:=d_i, \ 
    \delta_{\rm t}\xi_i:=\delta_i, \ 
    \delta_{\tau}t:=\tau, \quad i=1,2
  \end{split}
  \label{eq:Galileit}
\end{equation}
where the infinitesimal parameters 
$\phi_{\rm b,f}$, $v_i$, $d_i$, $\delta_i$ and $\tau$ are 
respectively the rotation angles of the bosonic and fermionic variables, 
the velocity of the Galilei boost of the bosonic variables and
the translation shifts of the
bosonic, fermionic and time variables. Among the parameters, 
only the $\delta_i$'s are Grassmannian, or fermionic.
The sLSZ model is shown to be strictly
invariant under the time and space
translations as well as  under the rotations,
\begin{equation}
  \delta_{\tau}L=0,\  \delta_{\rm t}L=0,\ \delta_{\rm r}L=0,
  \label{eq:Galileii}
\end{equation}
and furthermore, to be invariant under the Galilei boosts 
for the bosonic coordinates up to a total time-derivative,
\begin{equation}
  \delta_{\rm G}L=\dfrac{\D\Lambda_{{\rm G}}}{\D t},\quad
  \Lambda_{\rm G}=v_i\bigl(mx_i-k\epsilon_{ij}\dot{x}_j\bigr).
  \label{eq:LambdaG}
\end{equation}
We should note that the Grassmannian variables do not transform under
Galilei boosts: $\delta_{\rm G}\xi_i=0$.

The sLSZ model is also observed to have conformal and
superconformal symmetries~\cite{dHV}. Consider the
infinitesimal dilations, conformal and superconformal 
transformations given by
\begin{equation}
  \begin{split}
    & \delta_{\rm d}x_i:=g_{\rm b}\Bigl(t\dot{x}_i-\dfrac{1}{2}x_i
    -\dfrac{2k}{m}\epsilon_{ij}\bigl(t\ddot{x}_j-\dfrac{1}{2}\dot{x}_j
    \bigr)\Bigr), \\ 
    & \delta_{\rm d}\xi_i:=g_{\rm f}\Bigl(t\dot{\xi}_i
    -\dfrac{2k}{m}\epsilon_{ij}t\ddot{\xi}_j\Bigr) , \\
    & \delta_{\rm c}x_i:=h_{\rm b}\Bigl(t^2\dot{x}_i-tx_i
    -\dfrac{2k}{m}\epsilon_{ij}\bigl(t^2\ddot{x}_j-t\dot{x}_j\bigr)
    \Bigr),  \\
    & \delta_{\rm c}\xi_i:=h_{\rm f}\Bigl(
    t^2\dot{\xi}_i-\dfrac{2k}{m}\epsilon_{ij}t^2\ddot{\xi}_j
    \Bigr),  \\
    & \delta_{\rm s}x_i:=\I\beta\Bigl(t\xi_i-\dfrac{2k}{m}\epsilon_{ij}
    \bigl(t\dot{\xi}_j-\dfrac{1}{2}\xi_j\bigr)\Bigr) , \\
    & \delta_{\rm s}\xi_i:=-m\beta\Bigl(t\dot{x}_i-x_i
    -\dfrac{2k}{m}\epsilon_{ij}\bigl(t\ddot{x}_j
    -\dfrac{1}{2}\dot{x}_j\bigr)\Bigr),
  \end{split}
  \label{eq:conformalt}
\end{equation}
where $g_{\rm b,f}$, $h_{\rm b,f}$ and the Grassmannian variable $\beta$ are 
infinitesimal parameters.  In each of these cases, the Lagrangian remains  
invariant up to a total time-derivative:
\begin{align}
  & \delta_{\rm d}L=\dfrac{\D \Lambda_{\rm d}}{\D t},
  & \Lambda_{\rm d}:= & g_{\rm b}
  \Bigl(\dfrac{m}{2}t\dot{x}_i^2-3k\epsilon_{ij}t\dot{x}_i\ddot{x}_j
  +\dfrac{2k^2}{m}\bigl(t\ddot{x}_i^2-t\dot{x}_i\dddot{x}_i
  -\dfrac{1}{2}\dot{x}_i\ddot{x}_i\bigr)\Bigr) \nonumber\\
  & & & + \I g_{\rm f}\Bigl(\dfrac{1}{2}t\xi_i\dot{\xi}_i
  +\dfrac{2k}{m}\epsilon_{ij}t\bigl(\dot{\xi}_i\dot{\xi}_j
  -\dfrac{1}{2}\xi_i\ddot{\xi}_j\bigr)
  +\dfrac{4k^2}{m^2}t\dot{\xi}_i\ddot{\xi}_i\Bigr), \nonumber\\
  & \delta_{\rm c}L=\dfrac{\D \Lambda_{\rm c}}{\D t},
  & \Lambda_{\rm c} := & h_{\rm b}
  \Bigl(\dfrac{m}{2}\bigl(t^2\dot{x}_i^2-x_i^2\bigr)
  -k\epsilon_{ij}\bigl(3t^2\dot{x}_i\ddot{x}_j-x_i\dot{x}_j\bigr) \nonumber\\
  & & & +\dfrac{2k^2}{m}\bigl(t^2\ddot{x}_i^2-t^2\dot{x}_i\dddot{x}_i
  -t\dot{x}_i\ddot{x}_i\bigr)\Bigr) \label{eq:conformali}\\
  & & & + \I h_{\rm f}\Bigl(\dfrac{1}{2}t^2\xi_i\dot{\xi}_i
  +\dfrac{2k}{m}\epsilon_{ij}t^2\bigl(\dot{\xi}_i\dot{\xi}_j
  -\dfrac{1}{2}\xi_i\ddot{\xi}_j\bigr)
  +\dfrac{4k^2}{m^2}t^2\dot{\xi}_i\ddot{\xi}_i\Bigr), \nonumber\\
  & \delta_{\rm s}L=\dfrac{\D \Lambda_{\rm s}}{\D t},
  & \Lambda_{\rm s} := & \I\beta
  \Bigl(\dfrac{m}{2}\bigl(t\dot{x}_i\xi_i+x_i\xi_i\bigr)
  +k\epsilon_{ij}\bigl(t\ddot{x}_i\xi_j-3t\dot{x}_i\dot{\xi}_j
  +\dfrac{1}{2}\dot{x}_i\xi_j\bigr) \nonumber\\
  & & & +\dfrac{2k^2}{m}\bigl(2t\ddot{x}_i\dot{\xi}_i-t\dot{x}_i\ddot{\xi}_i
  -\dfrac{1}{2}\dot{x}_i\dot{\xi}_i\bigr)\Bigr). \nonumber
\end{align}
\section{The equations of motion and the canonical structure}
\label{sec:eom}
The Euler--Lagrange equations
\[
  \dfrac{\D}{\D t}\Bigl(\dfrac{\partial L}{\partial \dot{x}_i}
  -\dfrac{\D}{\D t}\Bigl(\dfrac{\partial L}{\partial\ddot{x}_i}\Bigr)\Bigr)
  -\dfrac{\partial L}{\partial x_i}=0 \text{ and }
  \dfrac{\D}{\D t}\Bigl(\dfrac{\partial L}{\partial\dot{\xi}_i}\Bigr)
  -\dfrac{\partial L}{\partial \xi_i}=0
\]
reduce to the following equations of motion for our model:
\begin{subequations}
  \begin{align}
    & \dfrac{\D}{\D t}\Bigl(
    m\dot{x}_i-2k\epsilon_{ij}\ddot{x}_j\Bigr)=0,\quad
    \dfrac{\D}{\D t}\Bigl(-\dfrac{1}{2}\I\xi_i
    +\dfrac{2k}{m}\I\epsilon_{ij}\dot{\xi}_j\Bigr)
    -\dfrac{1}{2}\I\dot{\xi}_i = 0,
    \label{eq:bELe} \\
    & \Leftrightarrow
    m\ddot{x}_i-2k\epsilon_{ij}\dddot{x}_j=0,
    \quad -\I\dot{\xi}_i+\dfrac{2k}{m}\I\epsilon_{ij}\ddot{\xi}_j=0.
    \label{eq:fELe}
  \end{align}
  \label{eq:ELe}
\end{subequations}\\
\noindent Note that the right derivative~\cite{C1,C2} is employed
to define the derivative in the fermionic coordinates.
This convention will be used throughout the paper.

Due to the presence of second order time-derivatives in the Lagrangian,
in order to formulate the sLSZ model in the Hamiltonian description of 
the Ostrogradski formalism, 
we need to introduce three kinds 
of momenta:
\begin{subequations}
  \begin{align}
    & p_i:=\dfrac{\partial L}{\partial \dot{x}_i}
    -\dfrac{\D}{\D t}\Bigl(\dfrac{\partial L}{\partial\ddot{x}_i}\Bigr)
    =m\dot{x}_i-2k\epsilon_{ij}\ddot{x}_j,
    \label{eq:p}\\
    & \tilde{p}_i:=\dfrac{\partial L}{\partial\ddot{x}_i}
    =k\epsilon_{ij}\dot{x}_j, 
    \label{eq:ptilde}\\
    & \pi_i:=\dfrac{\partial L}{\partial \dot{\xi}_i}
    =-\dfrac{1}{2}\I\xi_i+\dfrac{2k}{m}\I\epsilon_{ij}\dot{\xi}_j.
    \label{eq:pi}
  \end{align}
  \label{eq:canonical_momenta} 
\end{subequations}\\
\noindent This suggests that twelve canonical
variables $\{x_i,\dot{x}_i,p_i,\tilde{p}_i;\xi_i,\pi_i\}$ should be employed.
However, the elements in this set of canonical variables are not independent,
as can be seen from eq.~\eref{eq:ptilde}, which leads to two constraints,
\begin{equation}
  \Phi_i:=\dot{x}_i+\dfrac{1}{k}\epsilon_{ij}\tilde{p}_j=0,
  \label{eq:constraint}
\end{equation}
of the second class~\cite{D}.  Therefore, any physical
quantity can be described in terms of only ten coordinates.
For instance, using the Legendre transformation, 
\begin{equation}
  \begin{split}
    H & := \dot{x}_i p_i+\ddot{x}_i\tilde{p}_i+\dot{\xi}_i\pi_i - L \\
    & = -\dfrac{m}{2k^2}\tilde{p}_i^2-\dfrac{1}{k}\epsilon_{ij}p_i\tilde{p}_j
    -\dfrac{m}{4k}\I\epsilon_{ij}\bigl(\pi_i+\dfrac{1}{2}\I\xi_i\bigr)
    \bigl(\pi_j+\dfrac{1}{2}\I\xi_j\bigr) \\
    & = H_{\rm b}+H_{\rm f},  \\
    H_{\rm b}&:=-\dfrac{m}{2k^2}\tilde{p}_i^2
    -\dfrac{1}{k}\epsilon_{ij}p_i\tilde{p}_j,\quad
    H_{\rm f}:=-\dfrac{m}{4k}\I\epsilon_{ij}
    \bigl(\pi_i+\dfrac{1}{2}\I\xi_i\bigr)
    \bigl(\pi_j+\dfrac{1}{2}\I\xi_j\bigr),
  \end{split}
  \label{eq:Hamiltonian}
\end{equation}
we obtain the Hamiltonian of the sLSZ model in terms of the ten coordinates
$\{x_i,p_i,\tilde{p}_i;\xi_i,\pi_i\}$.

When
investigating the canonical equations of motion and the Poisson algebra
of the sLSZ model, it is necessary to use the graded Poisson bracket as well as the Dirac bracket.
Let $A,B$ be either bosonic or fermionic valued
differentiable functions of the canonical variables $\{x_i,\dot{x}_i,p_i,\tilde{p}_i;\xi_i,\pi_i\}$.
The graded Poisson bracket $\{A,B\}$ can be defined (in a non-graded form) as
\begin{equation}
  \{A,B\}:=
  \Bigl(
    \dfrac{\partial A}{\partial x_i} \dfrac{\partial B}{\partial p_i}
    -\dfrac{\partial A}{\partial p_i} \dfrac{\partial B}{\partial x_i}
  \Bigr)
  +\Bigl(
    \dfrac{\partial A}{\partial \dot{x}_i}
    \dfrac{\partial B}{\partial \tilde{p}_i}
    -\dfrac{\partial A}{\partial \tilde{p}_i}    
    \dfrac{\partial B}{\partial \dot{x}_i}
  \Bigr)
  -\Bigl(
    \dfrac{\partial B}{\partial \pi_i}
    \dfrac{\partial A}{\partial \xi_i}
    +\dfrac{\partial B}{\partial \xi_i}
    \dfrac{\partial A}{\partial \pi_i}
  \Bigr).
  \label{eq:gPB}
\end{equation}
The canonical Poisson brackets among the canonical variables are then
\[
  \{x_i,p_j\}=\delta_{ij},\ 
  \{\dot{x}_i,\tilde{p}_j\}=\delta_{ij},\ 
  \{\xi_i,\pi_j\}=-\delta_{ij},\ \text{others}\ \{\cdot,\cdot\}=0.
\]
Due to the constraints $\Phi_i$, we need to use the Poisson bracket
defined on the reduced phase space, which is nothing but the 
Dirac bracket~\cite{D},
\[
  \{A,B\}_{\rm D}:=\{A,B\}-\{A,\Phi_i\}C_{ij}\{\Phi_j,B\},
\]
where the matrix $C$ is defined through the relation
$C_{ik}\{\Phi_k,\Phi_j\}=\delta_{ij}$.
Substitution of the constraints \eref{eq:constraint} gives the Dirac
bracket for the sLSZ model:
\begin{equation}
  \{A,B\}_{\rm D}:=\{A,B\}-\{A,\Phi_i\}\dfrac{k}{2}\epsilon_{ij}\{\Phi_j,B\}.
  \label{eq:DiracB}
\end{equation}
Choosing the independent variables as 
$y_{a}:=\{x_i,p_i,\tilde{p}_i;\xi_i,\pi_i\}$, $a=1,\dots,10$,
we then have
\begin{equation}
  \{y_a,y_b\}_{\rm D}=\omega_{ab},
  \ 
  \omega:=\left[\begin{array}{ccccc}
   0 & E & 0 & 0 & 0 \\
   -E & 0 & 0 & 0 & 0 \\
   0 & 0 & \dfrac{k}{2}\epsilon & 0 & 0 \\
   0 & 0 &  0 & 0 & -E \\
   0 & 0 &  0 & -E & 0
  \end{array}\right],
  \label{eq:CDB}
\end{equation}
where
\[
  E:=\left[\begin{array}{cc} 1 & 0 \\ 0 & 1 
  \end{array}\right],\ 
  \epsilon:=\left[\begin{array}{cc} 0 & 1 \\ -1 & 0 
  \end{array}\right],
\]
and where $0$ denotes the $2\times 2$ null matrix.

Using the Dirac bracket, the canonical equations of motion read as
\[
  \dot{y}_a=\{y_a,H\}_{\rm D}+\dfrac{\partial y_a}{\partial t} .
\]
In the case of the sLSZ model, this leads to  
\begin{equation}
  \begin{split}
    & \dot{x}_i=-\dfrac{1}{k}\epsilon_{ij}\tilde{p}_j,\ 
    \dot{p}_i=0,\ 
    \dot{\tilde{p}}_i=-\dfrac{m}{2k}\epsilon_{ij}\tilde{p}_j
    -\dfrac{1}{2}p_i,\\
    & \dot{\xi}_i=\dfrac{m}{2k}\I\epsilon_{ij}\bigl(\pi_j
    +\dfrac{1}{2}\I\xi_j\Bigr),\ 
    \dot{\pi}_i=-\dfrac{m}{4k}\epsilon_{ij}\bigl(\pi_j
    +\dfrac{1}{2}\I\xi_j\Bigr),
  \end{split}
  \label{eq:cem}
\end{equation}
which is consistent with the Euler--Lagrange equations \eref{eq:fELe}
derived from the Lagrangian. We should note that the equations
of motion and the Dirac brackets of the Grassmannian variables 
can be cast into a simpler form
using the variables
\begin{equation}
  \theta_i^\pm:=\pi_i\pm\dfrac{1}{2}\I\xi_i,
  \label{eq:def_theta}
\end{equation}
as they then read
\begin{equation}
  \dot{\theta}_i^+=-\dfrac{m}{2k}\epsilon_{ij}\theta_j^+,\ 
  \dot{\theta}_i^-=0,\ 
  \{\theta_i^\pm,\theta_j^\pm\}=\mp\I\delta_{ij},\ 
  \{\theta_i^+,\theta_j^-\}=0.
  \label{eq:theta}
\end{equation}
We shall now investigate the Poisson algebra of 
the conserved charges of the sLSZ model.

\section{The Noether charges and their Poisson algebra}
\label{sec:lpa}

Let the sLSZ Lagrangian with its independent variables
$L=L(x_i,\dot{x}_i,\ddot{x}_i,\xi_i,\dot{\xi}_i)$ be denoted for short as
$L(x_i,\xi_i)$.  According to Noether's theorem, 
the invariance, up to a total derivative, of the Lagrangian $L$ with respect 
to the infinitesimal transformation, $\delta x_i$, $\delta\xi_i$, that is
\[
  \delta L := L(x_i+\delta x_i,\xi_i+\delta\xi_i)-L(x_i,\xi_i)
  =\dfrac{\D\Lambda}{\D t} ,
\]
implies the conservation of a quantity of the form
\begin{equation}
  C:=\delta x_i p_i+\delta\dot{x}_i \tilde{p}_i+\delta\xi_i\pi_i-\Lambda.
  \label{eq:NC1}
\end{equation}
Applying the formula \eref{eq:NC1} to
each symmetry transformations~\eref{eq:supert}, \eref{eq:Galileit} and
\eref{eq:conformalt} (except the time-translation) provides 
the following 12 conserved quantities:
\begin{align}
  &\text{space-translation:} & & \nonumber \\
  & C_{\rm t}= d_ip_i+\delta_i\theta_i^-,
  & & p_i,\ \theta_i^-,\nonumber\\
  &\text{rotation:} & & \nonumber\\
  & C_{\rm r}=  \phi_{\rm b}\left(\epsilon_{ij}x_i p_j
  -\dfrac{1}{k}\tilde{p}_j^2 \right)+\phi_{\rm f}\epsilon_{ij}\xi_i\pi_j,
  & & J_{\rm b}:=\epsilon_{ij}x_i p_j-\dfrac{1}{k}\tilde{p}_j^2, \ 
  J_{\rm f}:=\epsilon_{ij}\xi_i\pi_j,\nonumber\\
  &\text{Galilei boost:} & & \nonumber \\
  & C_{\rm G}= v_i(tp_i-mx_i+2\tilde{p}_i),
  & & G_i:=tp_i-mx_i+2\tilde{p}_i, \nonumber\\
  & \text{supersymmetric:} & & \nonumber \\
  & C_{\rm Q}=\alpha\bigl(p_i(\theta_i^+-\theta_i^-)
  -\dfrac{m}{k}\epsilon_{ij}\tilde{p}_i\theta_j^+\bigr),
  & & Q:=p_i(\theta_i^+-\theta_i^-)
  -\dfrac{m}{k}\epsilon_{ij}\tilde{p}_i\theta_j^+, \nonumber\\
  & \text{dilation:} & & \nonumber \\
  & C_{\rm d}=g_{\rm b}\Bigl(\dfrac{1}{2m}p_iG_i\Bigr),
  & & D:=\dfrac{1}{2m}p_iG_i, \label{eq:NCSLSZ1}\\
  & \text{conformal:} & & \nonumber \\
  & C_{\rm c}=h_{\rm b}\Bigl(\dfrac{1}{2m}G_i^2\Bigr),
  & & K:=\dfrac{1}{2m}G_i^2, \nonumber \\
  & \text{superconformal:} & & \nonumber \\
  & C_{\rm s}=\beta\Bigl(\dfrac{k}{m}\tilde{Q}-G_i\theta_i^{-}\Bigr),
  & & S:=\dfrac{k}{m}\tilde{Q}-G_i\theta_i^{-},\nonumber\\
  & & & \tilde{Q}:= \epsilon_{ij}p_i
  (\theta_j^+-\theta_j^-)+\dfrac{m}{k}\tilde{p}_i\theta_i^+. \nonumber
\end{align}
In order to replace the time-derivatives of the coordinates with 
the canonical momenta,
we have used the following relations derived from the definitions of the
canonical momenta~\eref{eq:canonical_momenta} and the equations of 
motion~\eref{eq:cem}:
\begin{align*}
  & \dot{x}_i=-\dfrac{1}{k}\epsilon_{ij}\tilde{p}_j,\ 
  \ddot{x}_i=\dfrac{1}{2k}\epsilon_{ij}p_j-\dfrac{m}{2k^2}\tilde{p}_i,\ 
  \dddot{x}_i=\dfrac{m^2}{4k^3}\epsilon_{ij}\tilde{p}_j
  +\dfrac{m}{4k^2}p_i,\\
  &  \dot{\xi}_i=\dfrac{m}{2k}\I\epsilon_{ij}\theta_j^+,\ 
  \ddot{\xi}_i=\dfrac{m^2}{4k^2}\I\theta_i^+.
\end{align*}
As can be seen in eq.~\eref{eq:SLSZ},
the Lagrangian of the sLSZ model does not explicitly depend 
on the time $t$, i.e. $\frac{\partial L}{\partial t}=0$.  Hence, 
the conserved quantity corresponding to the time-translation is given by
the Hamiltonian~\eref{eq:Hamiltonian}.

We have thus obtained 13 Noether charges:  
the Hamiltonian~\eref{eq:Hamiltonian} 
plus the 12 quantities appearing in \eref{eq:NCSLSZ1}.
We now turn to the Poisson algebra that these Noether charges generate.

Since the bosonic and the fermionic coordinates are decoupled in 
the Hamiltonian~\eref{eq:Hamiltonian}, the bosonic and fermionic parts
of the Hamiltonian, $H_{\rm b}$ and $H_{\rm f}$, are 
independently conserved.
Moreover, the canonical momenta $p_i$ are conserved, as is the
case in the interaction-free model.  We can thus separate 
the Hamiltonian~\eref{eq:Hamiltonian} into three individually conserved 
quantities, $H_0$, $H_k$ and $H_{\rm f}$,
\begin{equation}
  \begin{split}
    & H = H_0+H_{k}+H_{\rm f}, \\ 
    & H_0 := \dfrac{1}{2m}p_i^2,\ 
    H_k:=H_{\rm b}-H_0=-\dfrac{m}{2k^2}\tilde{P}_i^2,\ 
    H_{\rm f}=-\dfrac{m}{4k}\I\epsilon_{ij}\theta_i^+\theta_j^+,
  \end{split}
  \label{eq:separate_Hamiltonian} 
\end{equation}
where the quantities
\begin{equation}
  \tilde{P}_i:=\dfrac{k}{m}p_i+\epsilon_{ij}\tilde{p}_j
  \label{eq:LSZNCP}
\end{equation}
are the non-commuting modified momenta introduced
in ref.~\cite{LSZ}.
From the definition of the canonical momenta~\eref{eq:canonical_momenta},
we have
\begin{equation}
  \tilde{p}_i=k\epsilon_{ij}\dot{x}_j=O(k),\ 
  \tilde{P}_i=-\dfrac{2k^2}{m}\epsilon_{ij}\ddot{x}_j=O(k^2),\ 
  \theta_i^+=\dfrac{2k}{m}\I\epsilon_{ij}\dot{\xi}_j=O(k), 
  \label{eq:OE1}
\end{equation}
and thus observe immediately that
$H_k$ and $H_{\rm f}$ vanish in the limit $k\rightarrow 0$.

The Noether charge associated to the superconformal transformation $S$
in eq.~\eref{eq:NCSLSZ1} can also be divided into two independently conserved
quantities, $\tilde{Q}$ and $F$,
\begin{equation}
  S:=\dfrac{k}{m}\tilde{Q}-F,\ F:=G_i\theta_i^-,
  \label{eq:separate_S}
\end{equation}
since $G_i$ and $\theta_i^-$ are themselves conserved.

In addition to $F$,
three kinds of ``quadratic'' conserved quantities,
\begin{equation}
  \tilde{F}:=\epsilon_{ij}G_i\theta_j^-,\ 
  E:=p_i\theta_i^-,\ \tilde{E}:=\epsilon_{ij}p_i\theta_j^-,
  \label{eq:CQSLSZ}
\end{equation}
arise from the closure of the Dirac brackets.
These constructions are similar to that of $F$, 
in the sense that they are all
products of the linear Noether charges of eq.~\eref{eq:NCSLSZ1}.

The  Dirac brackets among the conserved charges,
\begin{equation}
  \{A,B\}_{\rm D}, \quad A, B\in\{p_i,\theta_i^{-},G_i,
  J_{\rm b}, J_{\rm f},Q,\tilde{Q},
  H_0,
  H_k, H_{\rm f},E,\tilde{E},D,F,\tilde{F},K\},
  \label{eq:current_algebra}
\end{equation}
are summarized in table~\ref{tb:current_algebra},
\begin{table}[ht]
\begin{center}
{\scriptsize
\begin{tabular}{c|m{3.8em}m{3.8em}m{3.8em}m{3.8em}m{3.8em}m{3.8em}m{3.8em}%
m{3.8em}}
 $A\backslash B$ & $p_j$ & $\theta_j^-$ & $G_j$ & $J_{\rm b}$ 
 & $J_{\rm f}$ & $Q$ & $\tilde{Q}$
 & $H_0$ \\ \hline 
 $p_i$ & 0 & 0 & $m\delta_{ij}$ 
 & $-\epsilon_{ij}p_j$ & 0  
 & 0 & 0 & 0 \\
 $\theta_i^-$ & & \underline{$\I\delta_{ij}$} 
 & 0 & 0 & \underline{$-\epsilon_{ij}\theta_j^-$} 
 & \underline{$-\I p_i$} & \underline{$\I\epsilon_{ij}p_j$} & 0 \\
 $G_i$ & & & $2k\epsilon_{ij}$ 
 & $-\epsilon_{ij}G_j$ & 0 & $m\theta_i^-$ 
 & $m\epsilon_{ij}\theta_j^-$ & $-p_i$ \\ 
 $J_{\rm b}$ & & & & 0 & 0 & $-\tilde{Q}$ & $Q$ 
 & 0 \\
 $J_{\rm f}$ & & & & & 0 & \underline{$-\tilde{Q}$} & \underline{$Q$} & 0 \\
 $Q$ & & & & & & \underline{$2\I mH$} & 0 & 0 \\
 $\tilde{Q}$ & & & & & & & \underline{$2\I mH$} 
 & 0 \\
 $H_0$ & & & & & & & & 0 
\end{tabular}\vspace{1ex}

\begin{tabular}{c|m{3.8em}m{3.8em}m{3.8em}m{3.8em}m{3.8em}m{3.8em}m{3.8em}%
m{3.8em}}
 $A\backslash B$ & $H_k$ & $H_{\rm f}$ & $E$ & $\tilde{E}$ 
 & $D$ & $F$ & $\tilde{F}$ & $K$ \\ \hline 
 $p_i$ & 0 & 0 & 0 & 0 & $\frac{1}{2}p_i$ 
 & $m\theta_i^-$ & $m\epsilon_{ij}\theta_j^-$ & $G_i$ \\
 $\theta_i^-$ & 0 & 0 & \underline{$\I p_i$} 
 & \underline{$-\I\epsilon_{ij}p_j$} 
 & 0 & \underline{$\I G_i$} & \underline{$-\I\epsilon_{ij}G_j$} & 0 \\
 $G_i$ & 0 & 0 & $-m\theta_i^-$ & $-m\epsilon_{ij}\theta_j^-$ 
 & ${\scriptstyle \frac{k\epsilon_{ij}}{m}p_j-\frac{1}{2}G_i}$ 
 & $2k\epsilon_{ij}\theta_j^-$ & $-2k\theta_i^-$ 
 & $\frac{2k}{m}\epsilon_{ij}G_j$ \\ 
 $J_{\rm b}$ & 0 & 0 & $-\tilde{E}$ & $E$ & 0 & $-\tilde{F}$ & $F$ & 0 \\
 $J_{\rm f}$ & 0 & 0 & \underline{$-\tilde{E}$} 
 & \underline{$E$} & 0 & \underline{$-\tilde{F}$} & \underline{$F$} & 0 \\
 $Q$ & $\frac{m}{2k}{\scriptstyle (\tilde{E}+\tilde{Q})}$ 
 & \underline{$-\frac{m}{2k}{\scriptstyle (\tilde{E}+\tilde{Q})}$}
 & \underline{$-2\I mH_0$} & 0 & $-\frac{1}{2}E$ & \underline{$-2\I mD$} 
 &\underline{${\scriptscriptstyle 2\I m\tilde{D}-4\I k H_{\rm f}^\dagger}$} 
 & $-F$ \\
 $\tilde{Q}$ & $-\frac{m}{2k}{\scriptstyle (E+Q)}$ 
 & \underline{$\frac{m}{2k}{\scriptstyle (E+Q)}$} & 0 
 & \underline{$-2\I mH_0$}
 & $-\frac{1}{2}\tilde{E}$ 
 & $\underline{\scriptscriptstyle 4\I kH_{\rm f}^\dagger-2\I m\tilde{D}}$ 
 & \underline{$-2\I mD$} & $-\tilde{F}$ \\
 $H_0$ & 0 & 0 & 0 & 0 & $H_0$ & $E$ & $\tilde{E}$ & $2D$ \\
 $H_k$ & 0 & 0 & 0 & 0 & 0 & 0 & 0 & 0 \\ 
 $H_{\rm f}$ &   & 0 & 0 & 0 & 0 & 0 & 0 & 0 \\
 $E$ & & & \underline{$2\I mH_0$} & 0 & $\frac{1}{2}E$ & \underline{$2\I mD$} 
 & $\underline{\scriptscriptstyle 4\I kH_{\rm f}^\dagger-2\I m\tilde{D}}$ 
 & $F$ \\
 $\tilde{E}$ & & & & \underline{$2\I mH_0$} & $\frac{1}{2}\tilde{E}$ 
 & $\underline{\scriptscriptstyle 2\I m\tilde{D}-4\I k H_{\rm f}^\dagger}$ 
 & \underline{$2\I mD$} & $\tilde{F}$ \\
 $D$ & & & & & 0 
 & ${\scriptstyle \frac{1}{2}F+\frac{k}{m}\tilde{E}}$ 
 & ${\scriptstyle \frac{1}{2}\tilde{F}-\frac{k}{m}E}$ 
 & ${\scriptstyle K+\frac{2k}{m}\tilde{D}}$  \\
 $F$ & & & & & 
 & ${\scriptscriptstyle\hspace*{-1em}%
 \underline{2\I mK+\frac{8\I k^2}{m}H_{\rm f}^\dagger}}$ 
 & 0 & $-\frac{2k}{m}\tilde{F}$ \\
 $\tilde{F}$ & & & & & &
 & ${\scriptscriptstyle\hspace*{-1em}
  \underline{2\I mK+\frac{8\I k^2}{m}H_{\rm f}^\dagger}}$
 & $\frac{2k}{m}F$ \\
 $K$ & & & & & & & & 0  
\end{tabular}
}
\end{center}
\caption{{\bf The Dirac brackets among the conserved charges.} 
{ The item in the $l$-th line and $c$-th column of the table 
is $\{A,B\}_{\rm D}$, where $A$ and $B$ are respectively the conserved
charges of the $l$-th line and $c$-th column.
We omit the items in the lower triangular part of the table since they can
be obtained, up to a sign, 
from the corresponding items in the symmetric position 
with respect to the diagonal (when the corresponding 
item is underlined, the signs
are the same.  Otherwise they are different).}}
\label{tb:current_algebra}
\end{table}
with $ \tilde{D}$ and $H_{\rm f}^\dagger$ standing in the table for:
\begin{align*}
  \tilde{D} & :=\dfrac{1}{2m}\epsilon_{ij}p_i G_j
  = \dfrac{1}{2}J_{\rm b}-\dfrac{k}{m}\bigl(H_0+H_k\bigr), \\
  H_{\rm f}^\dagger & := -\dfrac{m}{4k}\I\epsilon_{ij}\theta_i^- \theta_j^-
  = H_{\rm f}-\dfrac{m}{2k}J_{\rm f}.
\end{align*}

We note that $H_k+H_{\rm f}$ has a vanishing Dirac bracket with every
charge when it is placed to the right in the bracket,
\[
  \{A,H_k+H_{\rm f}\}_{\rm D}=0,\quad A\in\{p_i,\theta_i^{-},G_i,
  J_{\rm b}, J_{\rm f},Q,\tilde{Q},H_0,
  H_k, H_{\rm f},E,\tilde{E},D,F,\tilde{F},K\}.
\]
This result might suggest that $H_k+H_{\rm f}$  
belongs, like the charges associated 
to the mass $m$ and to the parameter
$k$, to the center of the Poisson algebra. But this is actually not the case, 
because it has a non-vanishing Dirac bracket with 
$Q$ and $\tilde{Q}$ when it is placed to the left of the bracket,
\[
  \{H_k+H_{\rm f},Q\}_{\rm D}=-\dfrac{m}{k}(\tilde{E}+\tilde{Q}),\ 
  \{H_k+H_{\rm f},\tilde{Q}\}_{\rm D}=\dfrac{m}{k}(E+Q),
\]
even though $\{H_k+H_{\rm f},\cdot\}_{\rm D}=0$ for every other bracket. 
In a similar way, we observe that
\[
  \{H_k-H_{\rm f},A\}=0,\ 
  \{Q,H_k-H_{\rm f}\}_{\rm D}=\dfrac{m}{k}(\tilde{E}+\tilde{Q}),\ 
  \{\tilde{Q},H_k-H_{\rm f}\}_{\rm D}=-\dfrac{m}{k}(E+Q),
\]
while $\{\cdot,H_k-H_{\rm f}\}_{\rm D}=0$, for all the other ones.
Such asymmetries in the Poisson algebra are also seen 
when charges made out of bosonic and fermionic parts are considered (like for instance 
$J_{\rm b}\pm J_{\rm f}$).

As we can readily see from 
eqs.~\eref{eq:separate_Hamiltonian} and  \eref{eq:OE1}, 
$H_k$ and $H_{\rm f}$
vanish in the limit $k\rightarrow 0$.
Even though the other conserved charges neither vanish nor diverge 
in this limit, some linear combinations of 
the charges accidentally vanish. This allows to reconcile our results
with those of the well-known $k=0$ situation.
For example, $E+Q$ and $\tilde{E}+\tilde{Q}$ that respectively appear 
as Dirac brackets of $\tilde{Q}$ and $Q$ with $H_k$ and $H_{\rm f}$ in
table~\ref{tb:current_algebra},
\begin{equation}
  \{Q,H_k\}_{\rm D}=\dfrac{m}{2k}(\tilde{E}+\tilde{Q})
  =-\{Q,H_{\rm f}\}_{\rm D},\ 
  \{\tilde{Q},H_k\}_{\rm D}=-\dfrac{m}{2k}(E+Q)
  =-\{\tilde{Q},H_{\rm f}\}_{\rm D},
  \label{eq:OE2}
\end{equation}
should vanish in the limit $k\rightarrow 0$ (even though 
a factor $1/k$ appears) since $H_k$ and $H_{\rm f}$ vanish 
in this limit.  Using the definitions of the 
charges~\eref{eq:NCSLSZ1} and \eref{eq:CQSLSZ}, 
as well as the expressions for the
canonical momenta and  the modified second momenta in the original
coordinates~\eref{eq:OE1}, one obtains that
\[
  E+Q=-\dfrac{4k^2}{m}\I\ddot{x}_i\dot{\xi}_i=O(k^2), \ 
  \tilde{E}+\tilde{Q}=-\dfrac{4k^2}{m}\I\epsilon_{ij}\ddot{x}_i\dot{\xi}_j
  =O(k^2),
\]
and thus that all Dirac brackets in eq.~\eref{eq:OE2} indeed vanish
when $k\rightarrow 0$. 
It should be remarked that both $E+Q$ and $\tilde{E}+\tilde{Q}$
have vanishing Dirac brackets with $\{p_i$, $\theta_i^-$, $G_i$, $H_0$,
$D$, $F$, $\tilde{F}$, $K\}$.

The above Poisson algebra~\eref{eq:current_algebra}
contains as subalgebras the
Galilei algebra, and the two subalgebras obtained by considering
only the quantities generated respectively by the bosonic and 
fermionic variables.
The linear generators
$\{p_i,\theta_i^+,G_i\}$ and the quadratic generators
$\{J_{\rm b}$, $J_{\rm f}$, $Q$, $\tilde{Q}$, $H_0$, 
$H_{k}$, $H_{\rm f}$, $E$, $\tilde{E}$, $D$, 
$F$, $\tilde{F}$, $K\}$ also
form subalgebras.  Besides such obvious subalgebras,
the Poisson algebra has a subalgebra generated by $\{p_i$, $\theta_i^-$,
$G_i$, $J_{\rm b}$, $J_{\rm f}$, $Q$, $\tilde{Q}$, $H_0$, $H_{k}$, 
$H_{\rm f}$, $E$, $\tilde{E}\}$,
which is the smallest subalgebra that simultaneously contains
the Noether charges associated with the Galilean and supersymmetric 
transformations as well as the full information of the Hamiltonian.
We should note that the interaction parts of the bosonic and fermionic
Hamiltonians, $H_{k}$ 
and $H_{\rm f}$, 
are in the centers 
of the subalgebras generated by the bosonic and fermionic variables 
respectively. We finally remark that there obviously exist 
other nontrivial subalgebras such as, for instance, 
$\{E+Q$, $\tilde{E}+\tilde{Q}$, $H_k+H_{\rm f}\}$.

\section{Concluding remarks}
\label{sec:remarks}

The sLSZ model \eref{eq:SLSZ} is a supersymmetric version of the 
Lagrangian~\eref{eq:LSZ}. As we have shown,
the sLSZ model remains invariant under the supersymmetric 
transformation~\eref{eq:supert},
the (2+1)-dimensional Galilean supersymmetry~\eref{eq:Galileit},
and the conformal and superconformal symmetries~\eref{eq:conformalt}.
Using the Ostrogradski--Dirac formalism 
for constrained Hamiltonian systems,
the Poisson algebra associated with the Noether charges of the
sLSZ model was investigated in detail, as was summarized in
table~\ref{tb:current_algebra}.

Conformal aspects of the bosonic model~\eref{eq:LSZ} are discussed
in ref.~\cite{SZ}, where the model is extended to include Coulomb and
magnetic vortex interactions. In the context of non-commutative
geometry, the non-commutative coordinates introduced in ref.~\cite{LSZ}
are modified as
\[
  X_i:=x_i-a\Bigl(\dfrac{2}{m}\tilde{p}_i
  -\dfrac{2k}{m^2}\epsilon_{ij}p_j\Bigr),
\]
where a dimensionless constant $a\neq 0$ can be chosen 
arbitrarily~\cite{HP},\footnote{Actually, the 
constant $a$ is fixed at unity in ref.~\cite{HP}.}
in order for the coordinates to behave as a Galilean vector,
\[
  \{G_i,X_j\}_{\rm D}=-t\delta_{ij}, \ 
  \{X_i,X_j\}_{\rm D}=-\dfrac{2k}{m}a^2\epsilon_{ij}.
\]
Other non-commutative coordinates introducing an interesting
split into ``external'' and ``internal'' degrees of freedom in
the sLSZ model are discussed in ref.~\cite{LSZ2}.
We expect further studies in such directions to be highly relevant
in the understanding of 
the spectrum generating algebra and the representation theory 
of the sLSZ model as well as its generalizations.

\section*{Acknowledgements} The authors are grateful to 
Professors P.~C.~Stichel and M.~S.~Plyushchay for 
their comments on this work.
One of the authors (HU) would like to express
his sincere gratitude to the Centre de Recherches Math\'ematiques,
the Universit\'e de Montr\'eal and McGill University for their 
warm hospitality.  This author is also supported by a 
grant for his research activities abroad from the Ministry of Education, 
Culture, Sports, Science and Technology
of Japan.  Part of this research was conducted while LL held a NSERC grant.
The work of LV is supported in part through a grant from NSERC.


\begin{thebibliography}{99}

\bibitem{SW} N.~Seiberg and E.~Witten, JHEP~{\bf 091999} 032 (1999).

\bibitem{Connes1} A.~Connes, Commun.~Math.~Phys.~{\bf 182} 155 (1996) 
[hep-th/9603053].

\bibitem{Connes2}A.~Connes, M.~R.~Douglas and A.~Schwarz,
JHEP {\bf 021998} 003 (1998).

\bibitem{Jackiw} See e.g., R.~Jackiw, {\it Lectures on Fluid Dynamics,
A Particle Theorist's View of Supersymmetric, Non-Abelian,
Noncommutative Fluid Mechanics and d-Branes},
CRM Series in Mathematical Physics, Springer, New York, 2002, Ch.~8.

\bibitem{DH1} C.~Duval and P.~A.~Horv\'athy, Phys.~Lett.~B~{\bf 479} 284 (2000).

\bibitem{LSZ} J.~Lukierski, P.~C.~Stichel and W.~J.~Zakrewski,
Ann.~Phys.~{\bf 260} 224 (1997).

\bibitem{DH2} C.~Duval and P.~A.~Horv\'athy, 
J.~Phys.~A:~Math.~Gen.~{\bf 34} 10097 (2001).

\bibitem{HP2} P.~A.~Horv\'athy and M.~S.~Plyushchay, 
Phys.~Lett.~B {\bf 595} 547 (2004) [hep-th/0404137].

\bibitem{LL} J.-M.~L\'evy-Leblond, {\it Galilei Group and Galilean Invariance},
in Group Theory and its Applications, Vol.~II (E.~M.~Loebl, Ed.) Academic
Press, New York, 1971.

\bibitem{LSZ2} J.~Lukierski, P.~C.~Stichel and W.~J.~Zakrewski,
{\it Noncommutative Planar Particles: Higher Order Versus First Order 
Formalism and Supersymmetrization}, in Group 24: Physical and Mathematical
Aspects of Symmetry, Proceedings of Group 24, XXIV-th International
Conference on Group-Theoretic Physics, Paris, July 15--21, 2002 
(J.P. Gazeau, R.~Kerner {\it et al.}, Ed.) IOP Publishing House, Bristol, 
2002 [hep-th/0210112].

\bibitem{C1} R.~Casalbuoni, Nuovo Cimento~{\bf 33A} 115 (1976).

\bibitem{C2} R.~Casalbuoni, Nuovo. Cimento~{\bf 33A} 389 (1976).

\bibitem{dHV} E.~D'Hoker and L.~Vinet, Commun.~Math.~Phys.~{\bf 97}
391 (1985).

\bibitem{D} P.~A.~M.~Dirac, {\it Lectures on Quantum Mechanics},
Dover Publications, 2001.

\bibitem{SZ} P.~C.~Stichel and W.~J.~Zakrewski, Ann. Phys.~{\bf 310} 158 (2004)
[hep-th/0309038].

\bibitem{HP} P.~A.~Horv\'athy and M.~S.~Plyushchay, JHEP {\bf 062002} 033 (2002)
[hep-th/0201228].

\end{thebibliography}
\end{document}